\documentclass[a4paper,conference]{IEEEtran}
\IEEEoverridecommandlockouts
\usepackage{cite}
\usepackage{amsmath,amssymb,amsfonts}
\usepackage{algorithmic}
\usepackage{graphicx}
\usepackage{textcomp}
\usepackage{xcolor}
\def\BibTeX{{\rm B\kern-.05em{\sc i\kern-.025em b}\kern-.08em
    T\kern-.1667em\lower.7ex\hbox{E}\kern-.125emX}}
\begin{document}

\title{An Experimental Study on the Functional Correctness of Network Device Configurations Produced by Large Language Models}

\author{\IEEEauthorblockN{Anastasios C. Politis}
\thanks{© 2026 IEEE. Personal use of this material is permitted. Permission from IEEE must be obtained for all other uses, in any current or future media, including reprinting/republishing this material for advertising or promotional purposes, creating new collective works, for resale or redistribution to servers or lists, or reuse of any copyrighted component of this work in other works. The final authenticated version is available at: https://doi.org/10.1109/IAICT71158.2026.11621016.}
\IEEEauthorblockA{\textit{Dept. of Computer, Informatics and Telecommunications Engineering} \\
\textit{International Hellenic University}\\
Serres, Greece \\
anpol@ihu.gr}
}

\maketitle

\begin{abstract}
This paper presents a preliminary study on the efficacy of network device configuration produced by Large Language Models (LLMs). Currently, the study focuses on routing protocol configurations in a predefined small internetwork with Cisco-based routing devices. The capability of three freely available LLMs with standard capabilities (i.e., with no enhanced features) to produce effective setups is investigated. Prompt engineering methods are excluded in order to assess the "first-response" configurations of the models under consideration. The assessment method is based on the contents of the resulting routing table, aiming at measuring the device functional correctness. For that purpose, the Jaccard index is used to indicate the percentage of correct network prefixes contained in the routing table constructed by the model-based configuration commands. The work provides insights on the maturity and limitations of modern LLMs on aiding network administrators with network device configurations in real deployments.
\end{abstract}


\section{Introduction}
Network devices form the infrastructure through which communication takes place. Moreover, they are the points in the network where the organization's operational intentions (e.g., the application of specific security policies) are applied. Hence, their proper and prompt configuration is critical in order to provide seamless services to end users and to implement organizational strategies. These devices may include routers, switches, wireless access points, firewalls etc. 

Traditionally, device configuration and parameterization was a manual process which was achieved through access to the Command Line Interface (CLI) of the appliance. Access to that interface is possible either locally or remotely\footnote{Remote access requires previous device configuration through local access for that purpose.}, where vendor-specific commands or configuration files can be imported to the operating system of the device. In more recent years, the Software Defined Network (SDN) technology allows operators to use protocols (e.g., Netconf and Restconf) and tools (e.g., Yang language) to simplify the configuration task through a central controller (e.g., OpenDaylight). However, in practical systems the presence of legacy and proprietary networking machinery, prohibits the full adoption of SDNs \cite{Yu2019, RFC7149}. Thus, human intervention in the configuration process through CLI is still frequently required \cite{ElHassany2017,Wei2025}. 

Manual configuration is considered a difficult task, which requires familiarity with vendor-specific software and documentation, often leading to erroneous system behaviour due to misconfigurations \cite{ElHassany2017,Wang2024}. To tackle the issue, several research works propose the use of configuration tools that translate intent-based languages into device-specific configuration commands \cite{Ramanathan2023,Tian2019}. Nevertheless, such approaches require familiarity with these high-level languages, adding an extra burden to the network administrating team.

LLMs have infiltrated the life of internet users, helping them in performing tasks such as researching, content creation and programming, by simply guiding the tool with the use of natural language. Hence, it is reasonable for one to question if these models are able to aid network device configuration, and thus, largely relieve network engineers from this cumbersome process. Unsurprisingly, the answer to that question is in the affirmative and naturally gives birth to a follow-up inquiry: \emph{is the network device configuration produced by LLMs sufficient to make the system operate as intended?}. 

Arguably, repetitively prompting an LLM with corrective input (a process known as prompt engineering) may lead to the desired outcome. However, such a procedure requires progressively greater human effort and introduces significant time overhead. Hence, we believe that a study on the ability of LLMs to produce device configurations, directly from their initial response (without exploiting prompt engineering methods), that lead to a functional networking system, is practically meaningful.

This paper performs a preliminary study on the functional correctness of network device configuration generated by general-purpose LLMs. The free version of three popular LLMs were specifically chosen to reflect the realistic setting of wide accessibility, without the leverage of any premium or enhanced capabilities. For the time being, the interest is restricted on routing protocol configurations. The LLMs chosen for the investigation include OpenAI's GPT, Google's Gemini and Mistral AI's Le Chat. Aiming at evaluating their capability to produce immediate functional configurations, prompt engineering approaches are avoided.

The rest of the paper is structured as follows. In Section II we review the works related to the current paper. Section III outlines the methodology followed and Section IV describes the experiments performed. Section V reports and discusses the outcome of the experiments and the paper is concluded in Section VI with final remarks and future directions.

\section{Related Works}

Exploiting LLMs for network device configuration is a relatively new field of research. This Section reviews some of the most recent and related papers on this field.

In \cite{Mondal2023}, the authors research the ability of GPT-4 to create correct router configurations with minimal manual intervention. They focus on using the LLM to translate Cisco router configuration into the corresponding Juniper format. Their findings reveal that the LLM's output configurations perform poorly containing several errors. To this end, they develop a system combining GPT-4 with a verifier that produces local feedback in order to automatically correct misconfigurations. 

The work in \cite{Naigong2024} partially utilizes LLMs to produce a unified device configuration translation framework. More specifically, LLMs are used to aid the translation of vendor-specific commands without parameters (e.g., commands that enable or disable certain functions, or to enter specific device views), while commands with parameters are translated based on a heuristic method. 

A similar work on device configuration translation exploiting LLM agents is explored in \cite{Wei2025}. A framework that uses LLMs to fragment a configuration file into pieces and extract intents, is proposed. Those intents are used for relevant manual extraction of the target device that is fed back to the LLM in order to create the appropriate configuration file. Syntax checkers and verifiers are also used to combat hallucinations. 

A part of the work presented in \cite{Wang2024} includes an experimental study closely related to the scope of our paper. More specifically, it investigates, among others, the effectiveness of routing configurations produced by GPT-4-Turbo. The LLM is used to output FRRouting settings by using prompt methods to update and extend the knowledge of the LLM. 

\section{Methodology}

The current work aims to evaluate the basic version of various LLMs in terms of achieving the desired device functionality. Versions with enhanced features typically require subscription, which may prohibit their adoption by network administrators. For this reason, the main requirement of our study is to utilize only the freely available LLM chatbots that are widely available to the broad user base and are lacking any advanced features.

A second constrain imposed in this work, is the avoidance of sophisticated prompt engineering techniques and continuous chatbot conversations to improve output configurations. In this way, the focus shifts from the user's competence to craft efficient prompts, to the tool's ability to produce unrefined, yet accurate, output configurations. Moreover, applying prompt engineering may be a time-consuming process, which, typically, is in contrast with the time-requirements of practical scenarios in device configuration. 

Lastly, given that lengthy prompts are proven to degrade the LLM's performance \cite{Liu2024}, we attempt to simulate a single and compact prompt, accompanied by the diagram of the network under consideration. An identical prompt is used across all LLMs included in this study. 

\begin{figure}[tb!]
\centering
\includegraphics[scale=1]{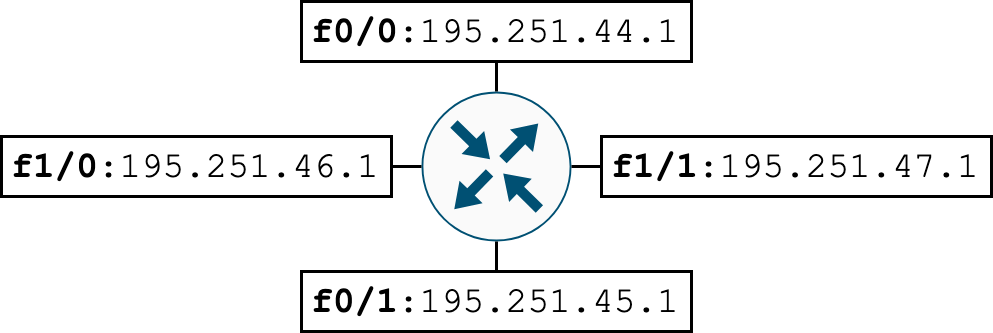}
\caption{An example of a routing device with four interfaces to be configured with OSPFv2 in the same area.}
\label{fig1}
\end{figure}

In terms of evaluation, many related works measure the similarity of low-level configuration commands produced by an LLM to a reference configuration command set and use the outcome as an assessment metric \cite{Wang2024,Wei2025}. Such a comparison can be misleading, as dissimilar configurations may lead to the desired device functionality. For example, consider the configuration procedure in multiple interfaces of the OSPFv2 protocol in the Cisco routing device shown in Fig \ref{fig1}. There are multiple command variations that can be used to reach the desired outcome. After entering the OSPF configuration mode with the command \texttt{router ospf 1}, the following four variations of subcommands can be used to enable the protocol in all interfaces:

\begin{figure*}[tb!]
\centering
\includegraphics[scale=0.8]{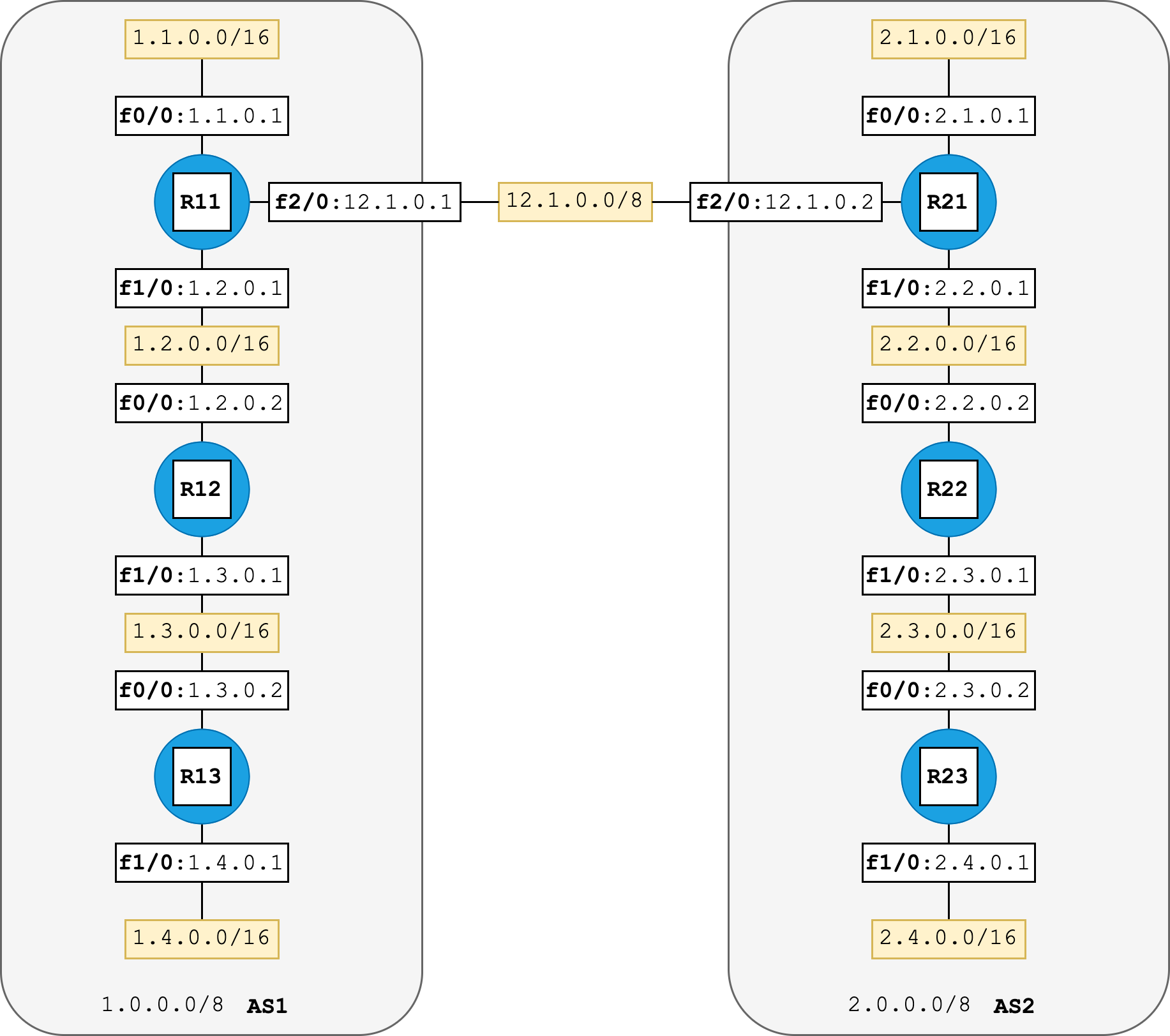}
\caption{The diagram for the internetwork used in the experiments.}
\label{fig2}
\end{figure*}

\noindent \texttt{network 195.251.0.0 0.0.255.255 area 0}

\noindent or

\noindent \texttt{network 195.0.0.0 0.255.255.255 area 0}

\noindent or

\noindent \texttt{network 195.0.0.0 255.255.255.255 area 0}

\noindent or

\noindent \texttt{network 195.251.44.1 0.0.0.0 area 0 \\ network 195.251.45.1 0.0.0.0 area 0 \\ network 195.251.46.1 0.0.0.0 area 0 \\ network 195.251.47.1 0.0.0.0 area 0}.

All of the above variants have the same effect: enable OSPFv2 in all interfaces of the device. Hence, a similarity check between the LLM's output and the reference commands may report lower matching score, even though the device performs as desired. For example applying the cosine text similarity method among the first and the last variant given above, a value of $\approx 0.55$ is produced (a value of $1$ indicates identical and $-1$ completely dissimilar texts). This outcome suggests fairly weak similarity, despite the fact that both command sets achieve the same goal.

Thus, another way to evaluate the output of the models was researched. Given that in this preliminary work we investigate the routing capability of the devices, the contents of the resulting routing table is chosen, which is compared to the reference routing table produced by the manual configuration process. The contents of the routing table reflect the reachability of remote networks, which fits the scope of measuring the device's desired operation.

To compare the contents of the reference and the LLM-produced routing tables, the Jaccard index is used. The Jaccard index measures the similarity/diversity of sample sets. Denoting with $P_r$ the set of network prefixes present in the reference routing table and with $P_l$ the prefix set contained in the respective table produced by the LLM under investigation, the Jaccard index is given by: 

\begin{equation}
J(P_r, P_l) = \frac{|P_r \cap P_l|}{|P_r \cup P_l|}.
\end{equation} 

We consider that the set of network prefixes inserted in the routing table, as consequence of applying the LLM's routing configuration commands, belongs to $P_r$ (i.e., $P_l \subseteq P_r$), and thus $P_r \cap P_l=P_l$ and $P_r \cup P_l=P_r$. Hence, the modified Jacard index used in the evaluation process, is expressed as:

\begin{equation}
J(P_r, P_l) = \frac{|P_l|}{|P_r|}.
\end{equation}

Obviously, $0\leq J(P_r, P_l) \leq 1$, with $J(P_r, P_l)=0$ meaning no common prefixes in the sets, and $J(P_r, P_l)=1$ indicating identical prefix sets.

\section{Experiments}

\begin{figure}[tb!]
\centering
\includegraphics[scale=0.78]{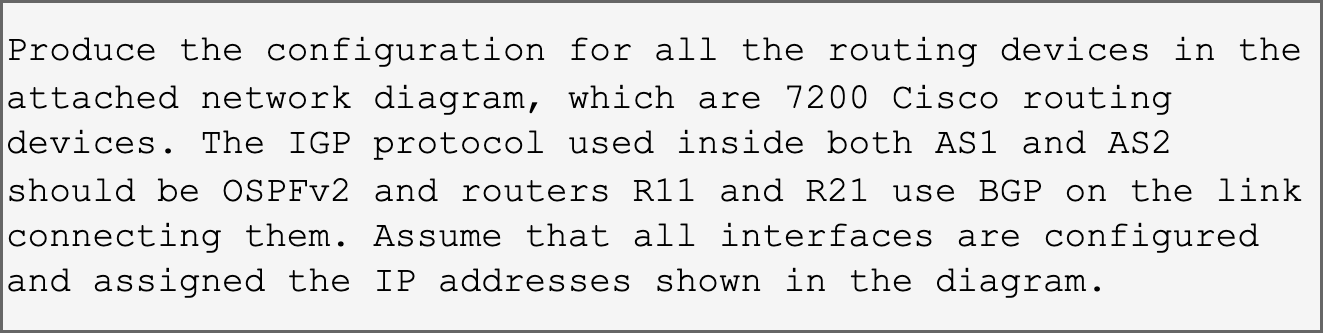}
\caption{The text used to prompt the models.}
\label{fig3}
\end{figure} 

\begin{figure*}[tb!]
\centering
\includegraphics[scale=0.7]{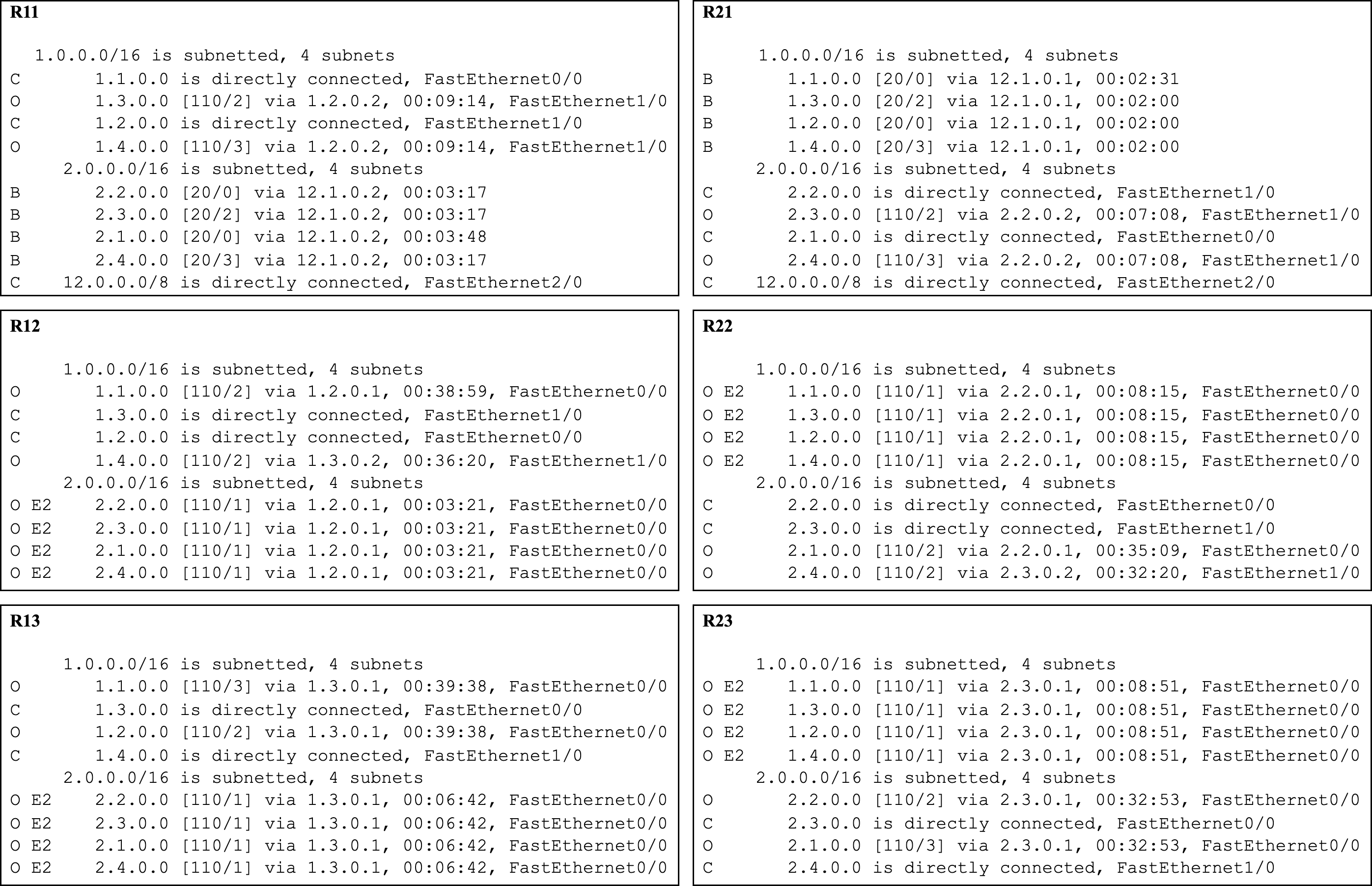}
\caption{Reference routing tables of all routing devices produced by manual configuration.}
\label{fig4}
\end{figure*}

The network diagram of the system under consideration is depicted in Fig. \ref{fig2}. The network is a variant of a network scenario involving several technologies from the Kathara project \cite{Kathara2018}. It involves two Autonomous Systems (ASs), \texttt{AS1} and \texttt{AS2}, with three routers each and several networks. The Interior Gateway Protocol (IGP) used in both ASs is the Open Shortest Path First (OSPF) protocol. Two routing devices, \texttt{R11} and \texttt{R21}, are used to connect both ASs by using the Border Gateway Protocol (BGP). Interface IDs and network/interface IP addresses are also clearly shown in the diagram. The network was initially built manually in the GNS3 emulator \cite{GNS3}, and all the appropriate commands are given to make the network fully functional as intended. The resulting routing tables from this process are used as a reference.

In an attempt to consider model diversity, the OpenAI's ChatGPT (GPT-5.3), the Google's Gemini (Gemini 3 Flash) and Mistral AI's LeChat (Mistral Medium 3) LLMs are chosen to be prompted for the device configurations. Exactly the same prompt was issued in all LLMs and included the diagram depicted in Fig. \ref{fig2}, as an attachment. The text input was concise, yet informative, and is depicted in Fig. \ref{fig3}.

After prompting each LLM, its response included the configuration per routing device. The AI generated configuration was inserted by simply copying and pasting the commands into the CLI of all routing devices.

\section{Experimental Results}

The routing tables of the routing devices produced by manual configuration, which are used as a reference, are shown in Fig. \ref{fig4}. It is visible that all intra and inter-AS networks are accessible by all routers. All entries, besides the ones representing directly connected networks (denoted with the code \texttt{C}), include an Administrative Distance (AD) and a cost value, denoted as \texttt{[AD,cost]}. \texttt{R11} and \texttt{R22}, can reach networks inside their AS via the OSPF protocol (denoted with the code \texttt{O}) and the networks to the remote AS via BGP (denoted with the code \texttt{B}). The cost for those remote networks is equal to the cost advertised by the peer BGP-enabled router that learned them via OSPF redistribution. The remaining routers can reach the networks on their remote AS which become known via BGP redistribution into OSPF (denoted with the code \texttt{O E2}). \texttt{E2} (External Type 2) indicates that the path is external to OSPF and its cost is not influenced by the internal OSPF metric calculation. The cost value equal to $1$ for these entries stems from the Cisco approach that during BGP redistribution configuration, if a metric is not specified, OSPF uses a default metric of $1$ \cite{cisco_redistribution}.  

\begin{table}[tbp]
	\caption{Jaccard index values for LLMs under consideration}
	\begin{center}
		\begin{tabular}{|c|c|c|c|}
			\hline
			\textbf{Device} &\textbf{ChatGPT} & \textbf{Gemini}&\textbf{LeChat} \\
			\hline
			\texttt{R11} & 1 & 1 & 0,16 \\\hline
			\texttt{R12} & 1 & 0,33 & 0,16 \\\hline
			\texttt{R13} & 1 & 0,33 & 0,16 \\\hline
			\texttt{R21} & 1 & 1& 0,16 \\\hline
			\texttt{R22} & 1 & 0,33 & 0,16 \\\hline
			\texttt{R23} & 1 & 0,33 & 0,16 \\\hline
		\end{tabular}
		\label{tab1}
	\end{center}
\end{table}

Table \ref{tab1} summarizes the Jaccard index for the LLMs investigated in this work. In the computation of the index, the entries for the directly connected network prefixes (code \texttt{C}) were excluded from the calculation process since their inclusion in the routing table depends solely on the correct IP configuration of the interfaces. This makes them independent of routing-oriented command sets.

Visual inspection of the contents of the resulting routing tables produced by the ChatGPT's generated configurations, revealed the total similarity with the contents of the reference tables, in terms of network prefixes. This led to a fully functional internetwork. ChatGPT's initial response with the device configurations were almost identical with the ones used during the manual configuration process. The noted differences included router ID configurations and a different command set on enabling OSPF in the desired interfaces. These discrepancies, however, did not contribute to malfunctioning network devices.

Gemini seems to provide effective configurations only for the \texttt{R11} and \texttt{R21} routers, while the remaining devices miss $66\%$ (i.e., four out of six) of the required network prefixes. More specifically all entries coded as \texttt{O E2} in the reference tables, were absent from the routing table created by the LLM's configuration commands. Inspecting the configuration commands produced by Gemini LLM, it was detected that in the autonomous system border routers (\texttt{R11} and \texttt{R21}), redistribution of OSPF learned prefixes towards BGP was configured, but not vice versa. Hence, a border router was not able to inject remote-AS network prefixes learned from its peer via BGP to the local OSPF process. This would allow them to propagate into the remaining devices of the local AS. 

The worst performance is exhibited by the LeChat LLM, whose generated configurations render the routing table of all devices incomplete with five out of six ($85\%$) network prefixes missing. Inspecting the configuration commands returned by the model, several misconfigurations were observed. Firstly, the interface \texttt{f2/0} in both \texttt{R11} and \texttt{R21}, which should run the BGP protocol, was also included in the OSPF configuration. This is a sign that the tool is suffering from hallucinations, which obviously contributed to the missing network prefixes in the routing tables. Moreover, redistribution commands were absent from both BGP and OSPF configuration modes. Lastly, in the BGP configuration part, the AS numbers were different from the ones included in the attached network diagram. This, however, was disregarded in the evaluation process of the model's performance. It must, also, be noted that, even though the prompt explicitly stated that interface IP addressing configuration should be considered complete, the LeChat LLM also returned a detailed command set to perform this task. This glitch was not considered in its evaluation. The same mishap was not observed in the generated configuration of the other two LLMs.

Overall, ChatGPT appears to have a high maturity level, exhibiting a trustworthy behavior in producing network device configuration commands with a single prompt. Gemini's outcome does not deviate significantly from the desired device functionality, requiring, however, user validation of its produced output. LeChat LLM seems to struggle with producing accurate network device configurations, leaving it in the last place of our investigation. 


\section{Conclusions}

In this work, a simple approach was followed in an attempt to investigate the correctness of routing configuration produced by freely available LLM chatbots. The LLM-proposed configurations were inserted into routing devices in a predefined network topology. The resulting contents of the routing tables were compared to a reference table which was created based in a manual configuration process in the GNS3 network emulator tool. We focused on three specific LLMs, namely OpenAI's ChatGPT, Google's Gemini and Mistral AI'a LeChat. Sophisticated prompt engineering techniques and perpetual chatbot conversations were avoided in order to assess the effectiveness of the LLM's initial response. Results indicate that ChatGPT is able to return fully effective configuration sets and Gemini LLM is closely following. LeChat exhibits the worst performance with incomplete generated routing tables due to the tool suffering from hallucinations. Future work will expand the study in assessing LLM-produced configurations for other services besides routing (e.g., Access Control Lists) and covering more LLMs in the investigation. Moreover, other network devices, such as switches and firewalls, are planned to be included.  

\bibliographystyle{IEEEtran}
\bibliography{references}

\end{document}